# Dual-comb dispersion measurement in LiNbO$_3$-on-insulator waveguides at telecom wavelengths

Halvor R. Fergestad,[1] Wolfgang Hänsel,[2] Arne Kordts,[2] Alessandro Prencipe,[1] Ronald Holzwarth,[2] and Katia Gallo[1]

[1]Nonlinear and Quantum Photonics group, Department of Applied Physics, KTH Royal Institute of Technology,

Roslagstullsbacken 21, Stockholm, SE-10691, Sweden

[2]Menlo Systems GmbH, Bunsenstraße 5, Martinsried, 82152, Germany

Photonic integrated circuits are a paramount platform for optoelectronics and nonlinear optics, enabling high confinement nonlinear interactions, where advanced waveguide dispersion engineering can be leveraged to tailor broadband nonlinear processes, supercontinuum generation, soliton formation and joint spectral amplitudes of photon pairs. Accurately measuring dispersion is therefore crucial for achieving efficient nonlinear and quantum devices. Here we develop a dual comb spectroscopy methodology suitable for dispersion measurements on photonic integrated circuits at telecom wavelengths and apply it to the characterization of lithium niobate on insulator ridge waveguides, mapping group velocity dispersion for fundamental TE and TM modes as a function of waveguide top-width. This work paves the way for the broader deployment of dual comb spectroscopy as a characterization tool for a broad range of photonic integrated circuit components on different nanophotonic platforms.

Dispersion engineering is an essential tool for optical applications, exemplified by dispersion compensating fibers deployed for the last 30 years in telecommunication systems to ensure data transfer over long-haul transmission systems [1], or more recently for ultrabroadband nonlinear devices, endless single-mode guidance or advanced spectral engineering with advanced fiber structuring [2-4]. In later years, photonic integrated circuits (PIC) have emerged as a powerful and versatile platform for the implementation of low-power, small-footprint, and ultrafast optical devices, leveraging the enhanced capabilities of high-confinement waveguides and opening new scenarios for dispersion engineering beyond the bounds traditionally imposed by material dispersion. With the added control knob of PIC waveguide geometry, a larger dispersion range is accessible and can be tailored according to the needs of specific applications. For instance, dispersion-compensating devices have been realized with PIC, reducing the footprint of fiberized systems as well as introducing dispersion tunability in fully integrated transceivers [5, 6]. Dispersion control is also relevant in anisotropic PIC material platforms, facilitating polarization division (de)multiplexing and polarization control in communication systems [7, 8]. Further applications have been rapidly emerging on PIC nanophotonic platforms, especially in the context of nonlinear and quantum on-chip devices. Here, dispersion engineering is a crucial element in designing and optimizing ultrafast operations [9], soliton generation [10], and entangled photon sources [11-13]. A considerable range of very compact nonlinear photonic devices has recently been developed in LiNbO$_3$ on insulator (LNOI), a novel PIC platform with enhanced nonlinear efficiencies, leveraged by $\chi^{(3)}$ and $\chi^{(2)}$ interactions [14, 15], and with unprecedented capabilities for dispersion engineering. Tailoring of group velocity dispersion (GVD) is, e.g. an essential tool for octave-spanning supercontinuum generation and for Kerr frequency combs at telecom wavelengths [16, 17]. Moreover, group velocity mismatch (GVM) engineering is opening up novel scenarios for broadband $\chi^{(2)}$ optical nonlinear interactions in periodically poled LNOI (PPLNOI) [18], with enticing perspectives currently addressing also the telecom band [9, 19]. Such an increasing number of applications coming of age sustain the quest for easy-to-use, versatile, and fast dispersion characterization tools, compatible with the PIC format and ideally enabling high-throughput device screening. The dispersion of on-chip integrated photonic structures can be measured in ring and disk resonator structures by tracking the frequency shift of their resonant lines [11, 20, 21]. However, this technique requires the fabrication of ad-hoc 2D resonators which has the intrinsic risk that the inferred dispersion is not fully representative for the target structure due to inhomogeneities over the substrate and, even worse, due to averaging over different propagation directions which is problematic for anisotropic samples [22]. In particular, the dispersion of very simple and fundamental structures such as straight waveguides is not easily assessed with the resonator approach. However, straight waveguides are especially interesting since they can be aligned with crystallographic axes over a long distance in order to support large nonlinear and/or electro-optic responses [9, 23]. It is therefore highly relevant to establish a method for in situ dispersion analysis of selected waveguide sections without the use of reference resonators. Traditional dispersion measurement techniques involve pulsed light probing of PIC and frequency resolved optical gating measurements [24], whose implementations in PIC formats is not always straightforward. As waveguide footprints shrink, the relevant phase changes for accurate dispersion retrieval may become prohibitively challenging to resolve. More generally, white-light interferometry can be used in various fashions. While the pure time-domain or spectral-domain measurements are better known, there are also combinations thereof. The authors of [25] use a tunable Michelson interferometer of a commercial Fourier spectrometer to analyze the signal from a Mach-Zehnder interferometer which encompasses the sample in one of its branches. This technique allows for spectrally resolved phase and amplitude retrieval. Here we make use of dual comb spectroscopy (DCS) to assess spectral phase in a similar fashion and apply it to dispersion measurements on PICs at telecom wavelengths. In particular, we investigate the dispersion of LNOI photonic chips, demonstrating the capabilities of the method even in the challenging conditions of strong polarization-dependence on relatively short straight waveguides. DCS is a maturing spectroscopic tool that combines many of the strengths of conventional broadband spectroscopy and tunable laser spectroscopy into a single platform [26, 27]. By using two coherent frequency combs and evaluating their spectral interference line-by-line, a Fourier spectrum is obtained without the use of moving parts. Even more, the use of frequency combs allows to relate the data to the SI second if required. With such a setup, one overcomes the response limitations and size constraints of conventional spectrometers. DCS has already proven its compatibility and value for spectroscopic measurements in gas-filled fibers [28], molecular line centers [29] and two-photon spectroscopy [30] as well as spectral lidar [31] and real-time precision ranging [32]. However, to the best of our knowledge, its potential for dispersion measurements of on-chip PIC waveguides has remained unexplored to date. The results of this Letter demonstrate the benefits of DCS in terms of fast detection, ease of implementation, and high spectral resolution in this new application area, paving the way for a broader deployment of DCS as a characterization tool for a large variety of PIC components on different nanophotonic platforms.

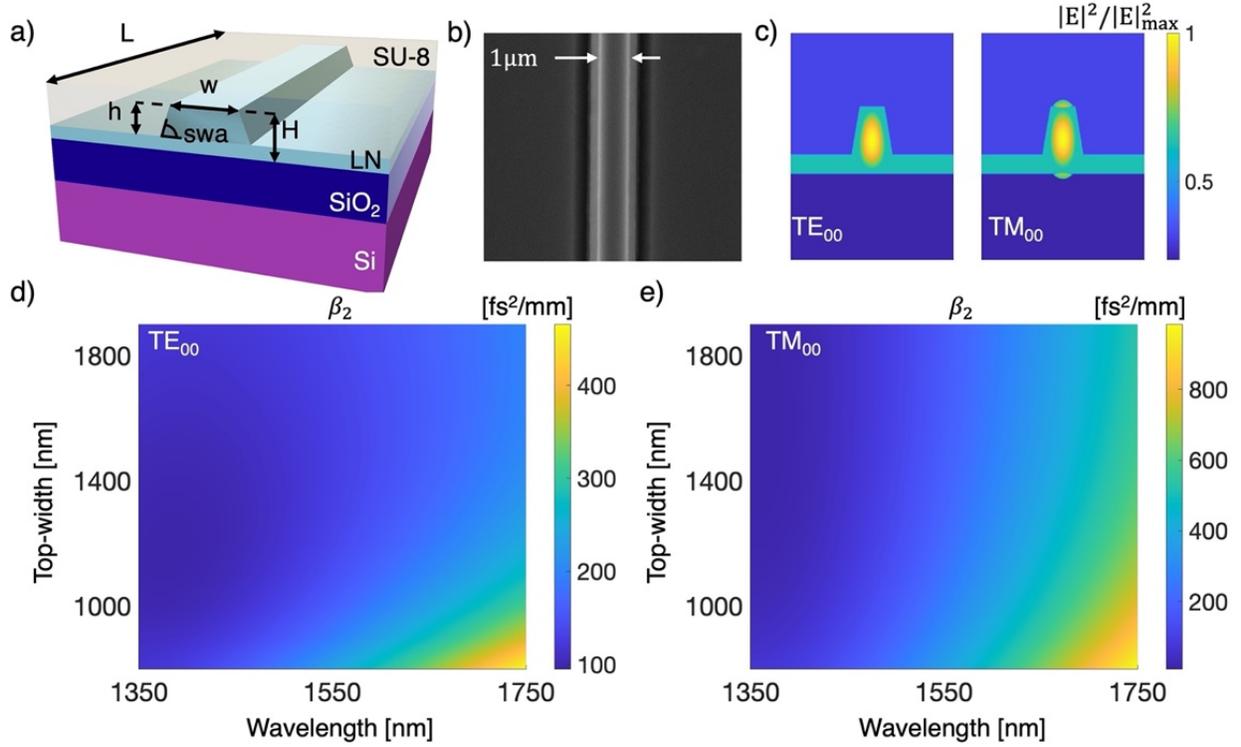

FIG. 1. a) Rendering of LNOI waveguide with cross-sectional parameters top-width (w), etch depth (h), lithium niobate thickness (H) and sidewall angle (swa), b) scanning electron microscope image of one of the LNOI waveguides encompassed in the dispersion study with w ≈ 1 μm, c) numerical field simulations of fundamental TE and TM modes in a LNOI waveguide with w ≈ 1 μm. d) and e) group velocity dispersion simulations for fundamental TE and TM modes, respectively, as a function of wavelength and waveguide top-width.

This study encompasses x-cut LNOI straight waveguides with varying waveguide top-widths, ranging from 800 to 1850 nm, considering fundamental modes in both polarizations (TE$_{00}$, TM$_{00}$) and their dispersion in the telecom C-band. The waveguides are etched 430 nm into a commercially acquired (NANOLN) LNOI with 600 nm LN film thickness. The sidewall angle after etching is estimated to 60 degrees through AFM/SEM inspection. The LN layer resides on a 500 μm thick silicon substrate of which the top face has been oxidized to form a 2 μm thick layer. The fabricated waveguide structure is protected by a ~10 um thick layer of SU-8 photo resist [33, 34]. Waveguides are patterned using electron-beam lithography and are subsequently etched using an Ar$^+$ reactive-ion etch recipe [23], followed by SU-8 cladding and end-facet preparation for butt-coupling with lensed fibers. Figure 1 shows the simulated GVD for the TE$_{00}$ (Fig. 1(b)) and TM$_{00}$ (Fig. 1(c)) waveguide modes as a function of wavelength and waveguide top-width, for the etch-depth of the fabricated LNOI sample. The simulations are performed using a finite difference eigenmode solver (Lumerical MODE). Due to the presence of SU-8 cladding, all waveguides are found to exhibit normal dispersion for both TE$_{00}$ and TM$_{00}$ modes. The GVD for the TE$_{00}$ (TM$_{00}$) mode ranges from 115 to 460 fs$^2$/mm (70 to 970 fs$^2$/mm) for an 800 nm-wide waveguide and from 120 to 200 fs$^2$/mm (12 to 520 fs$^2$/mm) in the largest (1850 nm-wide) one, when considering the 1350 – 1750 nm spectral interval.

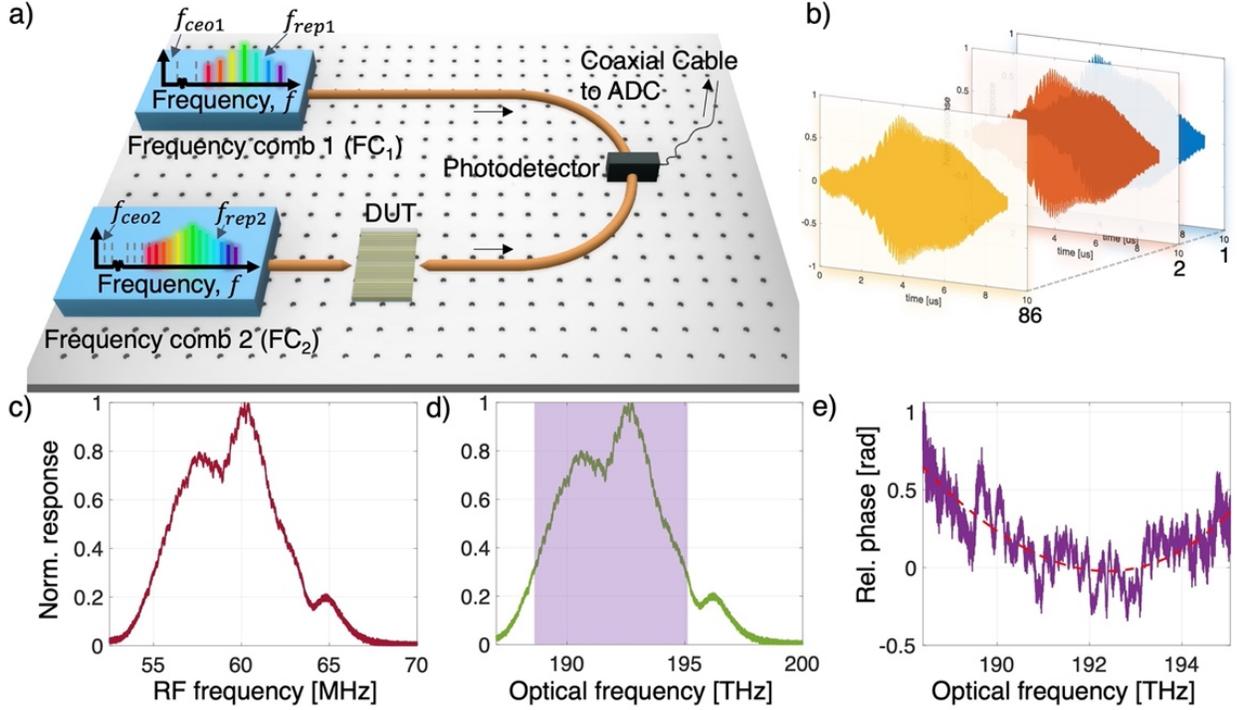

FIG. 2. Dual comb dispersion measurements on LNOI waveguides (DUT). a) Sketch of the experimental setup including two ultra-low noise femtosecond fiber laser frequency combs (FC$_1$ and FC$_2$), polarization maintaining fibers interconnecting the DUT and the two frequency combs to obtain a heterodyne signal on a photodetector, b) set of interferograms recorded by the A/D oscilloscope, recorded at a rate of 3.1 ms. The interferograms are coherently averaged and Fourier transformed for frequency analysis., c) Fast Fourier transform of 86 averaged interferograms, showing the RF spectral response centered around 62.5 MHz. d) Optical spectrum reconstructed as explained in the text. The purple shaded area indicates the spectral region used for phase retrieval, e) relative phase information retrieved from the experiment (purple) with corresponding quadratic fit (linear phase contribution is subtracted).

The experimental setup is shown in Figure 2 and uses two ultra-low noise femtosecond fiber laser frequency combs (FC$_1$ and FC$_2$) with sub-millihertz individual linewidth (FC-1500-ULN from Menlo Systems, wavelength ~1550 nm, repetition rate $f_{rep}$~250 MHz). Both frequency combs have their carrier-offset frequency locked via self-referencing [35, 36] and one comb mode optically locked to a shared reference laser at ~1542 nm (free-running with a frequency drift of ~100 MHz over 24 h). The selected stabilized modes m$_1$ and m$_2$ are placed at ~62.5 MHz from one another, which places the main components of the detected RF spectrum in proximity to this value. The repetition rate difference is chosen as $\Delta f_{rep} = f_{rep,2} - f_{rep,1} \approx 321$ Hz [37], thus defining the recurrence interval of the interferograms to 3.1 ms. FC$_1$ is used as a reference comb and FC$_2$ probes the device under test (DUT). The major part of interconnecting fibers is polarization-maintaining (PM), with only the short fiber leads to the DUT being non-PM and hence equipped with fiber polarization controllers (FPC) at input and output. The first FPC ensures probing the correct polarization state of the DUT while the second FPC ensures maximum heterodyne signal of the two combs on the photodiode. The A/D oscilloscope (ADQ214 from Teledyne - SP Devices) collects the 100 MHz low-passed heterodyne signal of the two pulse trains with a vertical resolution of 14 bits and a time resolution of 4 ns. The obtained interferogram thus consists of frequency components from the beat of the closest pairs of comb lines from the two frequency combs. In our case, the individual beat notes are located at $\Omega_{rf} = f_{m_2} - f_{m_1} + m \Delta f_{rep}$, with $f_{m_2} - f_{m_1} = 62.5$ MHz. The FPGA can continuously store up to 86 interferograms, corresponding to a measurement interval of 0.27 s. The data is read out by a computer which co-adds and performs averaging and the Fast Fourier transform, as illustrated in Fig. 2(c). The optical spectra, as depicted in Fig. 2(d), can directly be reconstructed by scaling the RF frequency axis to the optical axis according to:

$$f_{opt,2} = f_{CEO,2} + f_{rep,2}\left[m_2 + \frac{\Omega_{rf} - (f_{m_2} - f_{m_1})}{\Delta f_{rep}}\right] = f_{m_2} + f_{rep,2}\left[\frac{\Omega_{rf} - (f_{m_2} - f_{m_1})}{\Delta f_{rep}}\right]. \quad (1)$$

where $f_{opt,2}$ is the optical frequency of the probe comb, $\Omega_{rf}$ is the frequency component of the RF spectrum, $m_2$ is the index of the mode of FC$_2$, which is stabilized to the reference laser, and $f_{m_1}, f_{m_2}$ are the optical frequencies of the respective stabilized modes of the two frequency combs. The phase of the RF spectrum corresponds to the phase difference between the two contributing optical comb lines and can be extracted from the complex-valued RF spectral amplitude. The measurement procedure, as described so far, is performed first for the setup without the DUT and then with the DUT. The phase component from the initial calibration (including fiber tips but without DUT) is subtracted from the DUT measurement. The relative phase $\varphi(f)$ retrieved from the experiments (Fig. 2(e)), is fitted to a third order Legendre polynomial, $l$, in a spectral bandwidth corresponding to $\lambda = 1540 – 1590$ nm (purple region Fig. 2(d)). Legendre polynomials offer the nice feature that consideration of higher order terms does not influence lower terms. The GVD coefficient, $\beta_2$, is inferred through the second derivative as: $\beta_2 = (\partial^2 \varphi / \partial \omega^2)/L$, where $L$ is the waveguide length and $\omega = 2\pi f$.

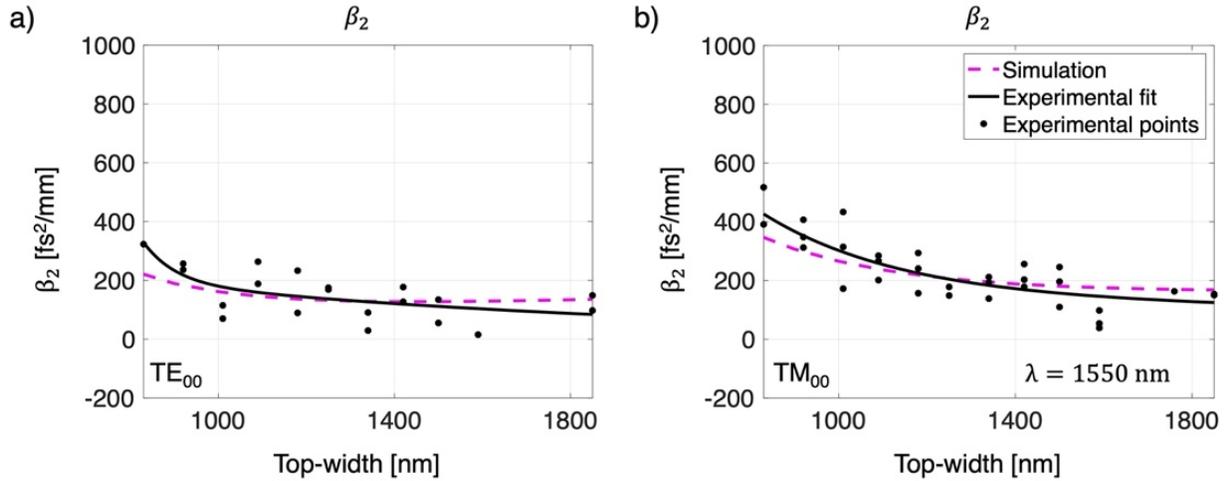

FIG. 3. DCS measured group velocity dispersion for a) TE and b) TM polarized modes in LNOI waveguides for wavelength 1550 nm as a function of waveguide top-width. Experimental points (black) are fitted with a dual-exponential fit (black line), simulation values (as in Fig. 1) are indicated as pink dashed line.

Figure 3 shows the results of $\beta_2$ measurements performed on TE$_{00}$ and TM$_{00}$ modes at 1550 nm in LNOI waveguides of length $L = 9$ mm and selected widths between 800 and 1850 nm. The experimental data points (circles) were fitted to a dual-exponential as a function of waveguide width, $w$ ($\beta_2(w) = Ae^{bw} + Ce^{dw}$, solid lines) and plotted alongside the GVD curves obtained by waveguide simulations using the waveguide nominal dimensions as per chip design (dashed lines). GVD for both polarizations is decreasing with increasing waveguide width, in good agreement with simulation. Experimental fit for TE$_{00}$ (TM$_{00}$) modes ranges from 330 (427) fs$^2$/mm for the 800 nm-wide waveguides to 84 (125) fs$^2$/mm for the widest 1850 nm waveguides. The experimental points range from 323 (517) to 15 (39) fs$^2$/mm, and the standard deviation between the experimental points and the experimental fit is 32 (35) fs$^2$/mm for TE$_{00}$ (TM$_{00}$) modes.

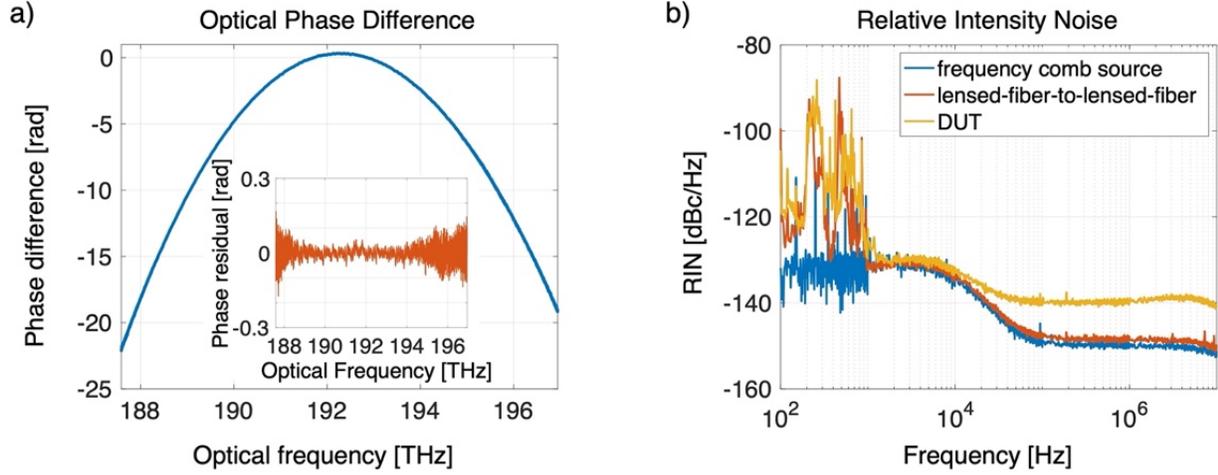

FIG. 4. a) Phase assessment of a 2 m long PM fiber with phase residual in inset. b) RIN measurements of the dispersion measurement setup in different configurations: 1) DCS setup with polarization maintaining fibers only, 2) DCS setup with lensed fibers but without the DUT, and 3) the DCS setup with lensed fibers and DUT.

While on a general scale there is good agreement between the experimental data and simulation, we observe considerable scatter on the data. The dual-comb method itself would suggest more reliable data. For comparison, fig. 4(a) illustrates the phase assessment of a 2 m long PM1550 fiber alongside with the fit residual, suggesting a far better phase estimate than observed in the data of Fig. 2(e). From the observed phase curvature, a group velocity dispersion of -23.7 $fs^2$/mm is inferred which is in agreement with previously reported chromatic dispersion values for PM1550 fiber [38]. A major contribution to the observed scatter in the waveguide assessment is suspected in acoustic noise which arises from vibrations in the DUT in- and out-coupling mechanics. This suspicion is corroborated by an investigation of intensity noise of the light transmitted through the DUT. The graph in Fig. 4(b) represents the relative intensity noise (RIN) measurements performed for different configurations of the DCS setup. The measurement is performed using a photodiode which is analyzed by an electronic spectrum analyzer. The RIN is measured for the frequency bandwidth 100 Hz (corresponding to a period three times as long as the acquisition time of one interferogram) to 10 MHz. The RIN of the frequency comb source is measured by substituting the two lensed fibers with a polarization-maintaining fiber connecting $FC_2$ directly to the photodiode. This yields the blue curve in Fig. 4(b), with RIN values decreasing from ~-130 dBc/Hz at 100 Hz to ~-150 dBc/Hz at 10 MHz. A direct comparison with the lensed fiber-to-lensed fiber configuration (red curve) highlights significant noise contributions appearing in the 100–1000 Hz bandwidth in the latter case, with a peak RIN of ~-90 dBc/Hz, hinting to noise sources connected to mechanical instabilities arising at the lensed fiber interfaces explaining the scatter that we observe on the data. When placing the DUT between the two lensed fibers (yellow curve in Fig. 4(b)), the RIN performance is similar to that without the DUT at low frequencies (100–1000 Hz), which clearly marks the mechanical mounting of the lensed fibers as cause of the instabilities. Due to lower signal level with the DUT, the noise floor is increased by ~10dB in that last measurement. Further investigations to identify the nature of all relevant noise sources, including those potentially arising from pyroelectric, nonlinear and photorefractive effects in LNOI chips are currently under way and will be the subject of a further publication. The data shown in Figure 4 allows to already identify directions for further improvements of the experimental setup to attain even higher accuracies in PIC dispersion measurements, which would benefit from stable pig-tailing solutions.

In conclusion, this work expands the application domain of dual comb spectroscopy to the burgeoning field of PIC devices, simplifying experimental procedures and affording direct measurements even in the challenging case of short, straight and highly birefringent waveguides, as demonstrated here with LNOI as a pilot. The method lends itself to an integrated, fast and easy-to-operate implementation, suitable for accurate dispersion analyses in the telecom C-band. We demonstrate its capabilities in polarization-resolved measurements of the group velocity dispersion as a function of waveguide width in high confinement x-cut LNOI photonic nanowaveguides. The dual-comb spectroscopy measurements were performed directly on the devices under test (9mm-long, straight waveguides on LNOI chips), with no need for additional assumptions, ad-hoc test structures such as ring resonators, spiral waveguides, or nonlinear interactions, to infer the linear dispersion properties of the waveguides. For 430 nm-high rib waveguides etched in 600 nm thick LNOI, with widths ranging from 800 to 1850 nm, corresponding to normal dispersion in the telecommunication C-band, we measure decreasing GVD values from 323 to 15 $fs^2$/mm for $TE_{00}$ modes and 517 to 39 $fs^2$/mm for $TM_{00}$ modes. These measurements are in good agreement with numerical simulations, highlighting the capability of the DCS setup for dispersion measurements. Moreover, we identify directions for further improvement of measurement resolutions with the elimination of noise sources. By reducing the RIN in our measurements higher order dispersion should be measurable, offering a wavelength resolved $β_2$ across the DCS bandwidth. Replacing the lensed fibers with objective lens coupling, although sacrificing in mode overlap with the DUT, appears as a promising route for improving the low frequency RIN (100 – 1000 Hz). The results are dense of implications for future developments of electro-optic, nonlinear optic, and quantum PICs and relevant, even more broadly, for on-chip dispersion and birefringence control in integrated nanophotonic architectures. Furthermore, they establish DCS as a fast and robust method of characterizing the dispersion of a platter of PIC components and platforms, even beyond the scope of the one specifically chosen here for a proof-of-principle demonstration of the concept.

## Acknowledgements

We acknowledge financial support from the European Union's Horizon 2020 Research and Innovation Program under the Marie Skłodowska-Curie grant agreement 812818 (MICROCOMB), from the Swedish Research Council (grants no. 2016-06122 and 2018-04487) and the Wallenberg Center for Quantum Technology (WACQT) in Sweden. The fabrication of the photonic integrated circuits has been carried out in the Albanova NanoLab facilities in Stockholm and the valuable technical support of its staff, particularly Erik Holmgren, Adrian Iovan and Taras Golod, is also gratefully acknowledged.

## Author Declaration

Halvor R. Fergestad, Alessandro Prencipe and Katia Gallo declare no conflict of interest. Wolfgang Hänsel, Arne Kordts and Ronald Holzwarth work at Menlo Systems GmbH.

## Author Contributions

**Halvor R. Fergestad:** Conceptualization (equal); Data curation (lead); Formal analysis (lead); Investigation (lead); Methodology (equal); Software (supporting); Validation (lead); Visualization (lead); Writing – original draft (lead); Writing – review & editing (equal).
**Wolfgang Hänsel:** Conceptualization (equal); Investigation (supporting); Software (lead); Supervision (supporting); Visualization (supporting); Writing – review & editing (equal).
**Arne Kordts:** Conceptualization (equal); Formal analysis (equal); Investigation (equal); Methodology (equal); Validation (supporting); Software (equal); Supervision (supporting); Writing – review & editing (supporting).

**Alessandro Prencipe:** Investigation (equal); Validation (supporting).

**Ronald Holzwarth:** Conceptualization (equal); Investigation (supporting); Validation (supporting); Resources (equal); Supervision (supporting); Writing – review & editing (supporting).

**Katia Gallo:** Conceptualization (supporting); Formal analysis (supporting); Funding acquisition (lead); Investigation (equal); Methodology (equal); Project administration (lead); Resources (lead); Supervision (lead); Validation (supporting); Writing – original draft (supporting); Writing – review & editing (equal).

## Data Availability

The data that support the findings of this study are available from the corresponding author upon reasonable request.